\documentclass[aps,prb,twocolumn,superscriptaddress,showpacs]{revtex4}

\usepackage{graphicx}
\usepackage{epsfig}

\def\gamnas{{Ga$_{1-x}$Mn$_{x}$As}}

\begin{document}


\title{Optimally Designed Digitally-Doped Mn:GaAs}



\author{J.~L. Xu}
\affiliation{Department of Chemical and Materials Engineering,
 Arizona State University, Tempe, AZ, 85287}
\email[]{Mark.vanSchilfgaarde@asu.edu}

\author{M. van Schilfgaarde}
\affiliation{Department of Chemical and Materials Engineering,
 Arizona State University, Tempe, AZ, 85287}

%

\date{\today}

\begin{abstract}

We use the {\em ab initio} local-density approximation (LDA) to
study of exchange interactions and $T_c$ of {\gamnas} grown in
digitally doped structures.  We analyze the crystallographic
dependence of exchange interactions predicted by the LDA in terms
of the Mn $t_2$ and $e$ levels, and explain the origin of the
antiferromagnetic contribution to the total exchange interactions.
We exploit this dependence and consider $\delta$-doping in specific
orientations where the antiferromagnetic interactions are minimized, to
optimize $T_c$ of the system.  By including hole doping with the addition
of Be in the GaAs host digitally doped {\gamnas} is predicted to be
significantly above room temperature.

\end{abstract}


\maketitle

Spintronics-based materials have attracted a lot of interest
recently \cite{Ohno96, Ohno99}.  By introducing spin as a degree
of freedom existing technologies can be improved, e.g. smaller
devices \cite{Datta90}

that consume less electricity, and are more powerful for certain
types of computations as compared to present charge-based
systems.  A key point is that the magnetism be carrier-mediated,
i.e. the magnetic state can be manipulated by electrical or
optical means.  Dilute magnetic semiconductors (DMS),
i.e. semiconductors doped with low concentrations of magnetic
impurities Mn, Cr, or Co, especially Mn doped GaAs, are generally
thought to be good candidates to meet the requirements for
spintronics applications.  A central issue that has impeded the
spintronics field is the lack of a (demonstrably)
carrier-mediated ferromagnetic semiconductor with critical
temperature $T_c$ above room temperature.

In \gamnas, Mn ions (approximately) randomly occupy the Ga sites
and act as both spin providers ($S$=5/2) and acceptors.  Because
ferromagnetic coupling between local ions is mediated by carriers
(holes in \gamnas), the doping concentration is believed to be a
crucial factor which determines the ferromagnetic properties of
DMS.

Traditionally, people use molecular beam epitaxy (MBE) methods to
grow DMS thin films with magnetic elements randomly doped into the
host semiconductors \cite{Ohno96,Chiba03,Ku03}.  However, this
preparation process always suffers of the low solubility of the
magnetic elements, and the tendency of Mn to form interstitials
deleterious to the magnetism.  Although annealing significantly
improves the quality of the samples, optimized Curie
temperatures ($T_c$) for these random alloys are below
200K\cite{Edmonds04}, well below requirements for room temperature
applications. Although $T_c$ is predicted to be enhanced when the
material is co-doped with acceptors, attempts to do this with
holes has not been successful.\cite{Wojtowicz03}

An alternative to a random alloy, $\delta$-doping of Mn in a GaAs
host has also been attempted with some success
\cite{Kawakami00,Nazmul03,Nazmul05,Wojtowicz03}.  A very thin
MnAs layer is periodically doped in a GaAs epitaxial layer,
produce high Mn doping concentration in a few (typically 2 to 3)
GaAs monolayers (ML).  Because the Mn-derived impurity band sits
slightly above the GaAs valence band, and is approximately
localized within the Mn layer, the holes are thus confined, which
according to simple models should increase $T_c$.
\cite{Sham01}.

In this paper, we use a first-principles linear response approach
based on the local-density approximation (LDA), to study
ferromagnetism in $\delta$-doped \gamnas, and contrast it to the
random alloy.  We show the following, and explain each point
qualitatively from the electronic structure of Mn $d$ levels: (i)
The magnetic exchange interactions have a strong crystallographic
and concentration dependence.  As a result: (ii) There is an
optimal Mn concentration within the $\delta$-layer; (iii) There is
a optimal $\delta$-layer thickness; (iv) $p$-type doping (Be in
GaAs) in the substrate can increase $T_c$ of the system by a large
amount, bringing it well above room temperature.

We adopt a linear-response approach\cite{licht87} valid in the
rigid-spin and long-wave approximation\cite{vladimir03}, in which
the change in energy from noncollinear spin configurations is
mapped onto a Heisenberg form $H=-\sum_{RR'}J_{RR'}\, {\hat
e_R}\cdot{\hat e_{R'}}$.

In multiple-scattering theory the $J_{RR'}$ are computed from the
scaled susceptibility\cite{licht87}:
\begin{equation}
J_{RR'} = \frac{1}{\pi} {\rm Im}\int^{\varepsilon_F} \sum_{LL'}
d\varepsilon\,\{ \delta P_{RL} T_{RL,R'L'} T_{R'L',RL} \delta P_{R'L'}
 \}
\label{eq:jij}
\end{equation}
The notation and computational approach is based on the Atomic
Spheres approximation and is described in detail in
Ref.~\onlinecite{mark99}.  $T_{RL,R'L'}$ is related to the
Green's function and $\delta P_{RL}$ is related to the exchange
splitting.  $L$ is a compound index referring to the $l$ and $m$
quantum numbers of the orbitals at a site.

Because of the large Mn local moment, the approximations are
expected to be reasonable.  A previous paper using the same
approach presents results for random alloys \cite{Xu05}. The
finite temperature magnetization $M(T)$ is obtained from the
Heisenberg form assuming classical spins.  The cluster variation
method (CVM)\cite{kikuchi} adapted to the heisenberg hamiltonian
was used to obtain $T_c$; it was shown in Ref~.\onlinecite{Xu05}
that the (CVM) does a rather good job of estimating $T_c$ in
random DMS alloys, and is expected to be comparably accurate here.

All the calculations are performed using supercells with periodic
boundary conditions.  The supercell we use includes 8ML in the
[001] direction.  We set the  thickness of Mn layer to be 1 or 2
ML, which is comparable with the experiment setup.  The
inter-layer distance is chosen to be 6 ML which is an optimized
value based on both speed and accuracy considerations.  Checks
show that further increase of the interlayer distance does not
significantly affect the exchange interactions $T_c$.  The lateral
scale of the layers are chosen based on Mn concentration
simulated.  Both ordered and Special Quasi-Random Structures
(SQS)\cite{zunger90} are used to simulate Mn doping inside the
$\delta$ layer.

\begin{figure}[ht]
\centering
\epsfig{file=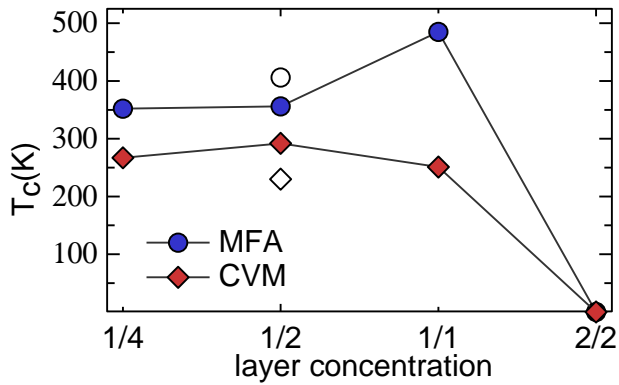,angle=0,width=0.45\textwidth,clip}
\caption{Dependence of $T_c$ on Mn concentration inside an
  superlattice consisting of a 1-2 ML thick $\delta$-layer of Mn,
  sandwiched by a 6 ML thick GaAs layer. Mn are taken to
  form an ordered structure within the $\delta$-layer.  The $x$
  axis refers to the fraction of Ga atoms that are replaced by Mn in
  the layer considered. In the ``$m/n$'' label along $x$ axis, $m$ represents the
 number of ML in the Mn doped layer, and $n$ is the total number of
cation sites in the doped layer. Circles show mean-field results
  for comparison. The 2/2 system is predicted to be a frustrated
  spin glass; this is reflected in the figure by assigning
  $T_c$=0.  Open diamond (circle) : $T_c$ for an SQS structure
  spread over 1ML at 50\% concentration calculated within the CVM
  (MFA).}
\label{fig:concentration}
\end{figure}



We first investigate a series of ordered $\delta$-layers varying
concentration within the layer for 1-2 ML superlattices.
Fig~\ref{fig:concentration} shows the calculated $T_c$ for the Mn
concentration in the $\delta$-layer ranging between 1/4 and 1.\cite{TwoDHeisennote}

For bulk random alloys\cite{Xu05}, the disorder in the Mn site
positions rather strongly affects exchange interactions and
reduces $T_c$.  In the bulk {\gamnas} alloy, the predicted optimum
$T_c$ was found to be $\sim$300K at $x$=12\%.
Fig.~\ref{fig:concentration} compares an ordered to a disordered
configuration of Mn within the $\delta$-layer in the 1/2 filling
case (Mn position in the ML were simulated by an SQS structure).
It is seen that the effect of disorder is to reduce $T_c$
slightly, just as in the bulk random alloy.  (Note that mean-field
theory incorrectly predicts $T_c$ {\em increases} with disorder,
as was also found in the bulk alloy.)

\begin{figure}[ht]
\epsfig{file=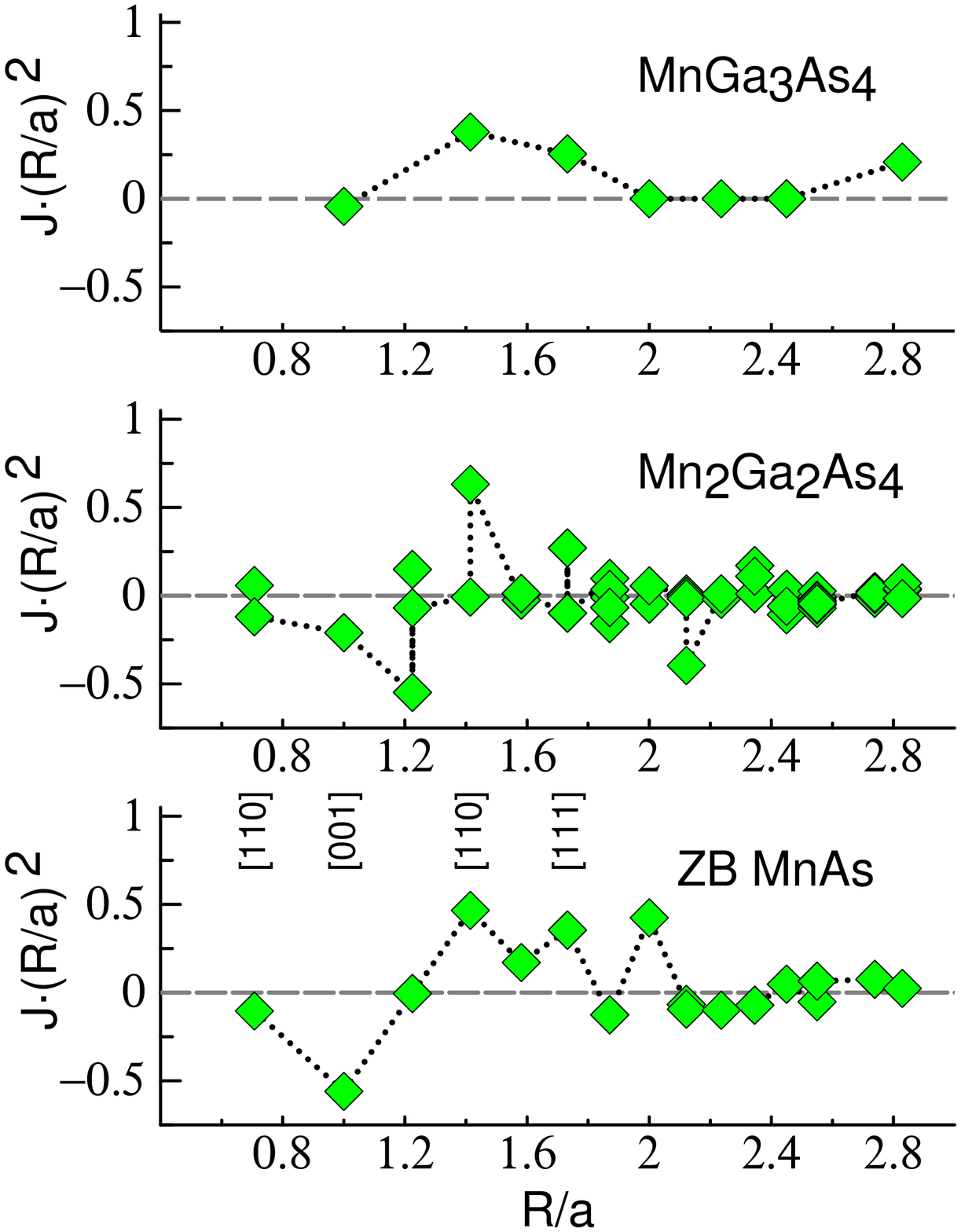,angle=0,width=0.45\textwidth,clip} \vskip
12pt
\begin{tabular}{
r@{\,\,\,}|@{\,\,\,}r@{\,\,\,}|@{\,\,\,}r@{\,\,\,}|@{\,\,\,}r@{\,\,\,}|@{\,\,\,}r@{\,\,\,}|}
  Structure  &   $t_2$  &  $e$      & $sp$  & total  \\
\colrule
  L1$_2$     &   470    &  -104     & 16    & 382    \\
  SQS-4      &   102    &  -250     & 76    & -71    \\
  ZB         &    -6    &  -206     & 78    & -134
\label{table:jlres}
\end{tabular}
\caption{Scaled exchange interactions $J_{RR'} \times R^2$ as a
  function of separation $R=|\bf{R}-\bf{R}'|$.  $R$ is in units
  of the lattice constant $a$. (top)
  MnGa$_3$As$_4$ in the L1$_2$ structure; (middle)
  the SQS-4 Mn$_2$Ga$_2$As$_4$; (bottom) MnAs in the ZB phase.
  In the SQS-4 structure a given $R$ be found in different environments,
  a rather strong dispersion in $J_{RR'}$ for a particular $R$ is
  evident.  The Table shows resolution of $T^{\rm MFA}_c$ into
  contributions from the Mn $t_2$, $e$ and $sp$ orbitals.
}
\label{fig:jij}
\end{figure}

Notably, the optimum $T_c$ for the $\delta$-layer case occurs near
50\% filling in a single ML and the optimum $T_c$ is near that
predicted for the bulk case ($\sim$300K at $x$=12\%). Most of this
effect can be traced to the strong crystallographic dependence of
the magnetic exchange interactions. This can be seen by
considering exchange interactions $J_{RR'}$ in {\gamnas} compounds
of high Mn concentration (Fig.~\ref{fig:jij}). While there is some
variation between compounds, there is a tendency for the $J_{RR'}$
to be \emph{weak or even} antiferromagnetic along the [001]
axis, but to be \emph{more strongly} ferromagnetic along the [110]
axis.  A similar result is found for the dilute random
alloys\cite{Xu05}. The origin of the strong crystallographic
dependence can, at least in part, be explained in terms of the
orbital character of the Mn $d$ orbitals of $t_{2}$ and $e$
symmetry.  Mn impurities in GaAs form a ``dangling bond hybrid''
level\cite{Mahadevan04} of $t_2$ symmetry which sits $\sim0.1$eV
above the valence band maximum, and is an admixture of the Mn $d$
levels of $t_2$ symmetry ($xy$, $yz$, and $xz$) and effective-mass
like states derived from the top of the host valence band.  Below
the valence band top sit Mn $d$ orbitals of $e$ symmetry
($3z^2-r^2$ and $x^2-y^2$), which hybridize less strongly with the
environment than the $t_2$ orbitals.  As is well known, the $e$
states are filled, while the three $t_2$ states are partially
filled with 2 electrons. (Mn, with 5 $d$ electrons and 2 $sp$
electrons, substitutes for Ga, with 3 $sp$ electrons.  Thus Mn
contributes 5 $d$ electrons and a hole.)

\begin{figure}[ht]
\centering
\epsfig{file=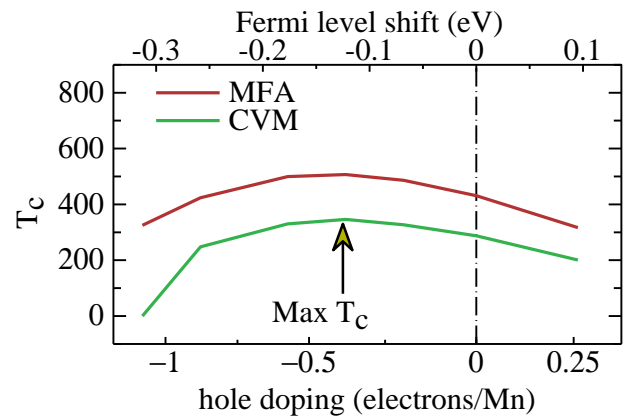,angle=0,width=0.45\textwidth,clip}
\caption{Dependence of $T_c$ on the Fermi level shift and excess
hole concentration
  in a bulk SQS Ga$_{92}$Mn$_{8}$As$_{100}$ alloy.  $E_F=0$
  corresponds to the usual charge-neutral case.}
\label{fig:tcfermi}
\end{figure}

While there is some disagreement as to whether the exchange is
best described by an RKKY-like picture\cite{dietl00} or a double
exchange picture\cite{akai98}, it is generally agreed that FM in
{\gamnas} derives from the partial filling of the $t_2$ level.
According to the Zener-Anderson-Hasegawa
model\cite{Zener51,Anderson55} of double exchange/superexchange,
exchange interactions between sites depend on the filling.  At
half filling the interactions are ferromagnetic, while filled
levels couple antiferromagnetically.  To the extent that this
model is a valid description of FM in the {\gamnas} system, the
exchange interactions should be maximally FM when the $t_2$ level
is half full.  This is apparently a reasonable description of the
LDA exchange interactions, as Fig.~\ref{fig:tcfermi} shows.
Exchange parameters $J_{RR'}$ were computed in a bulk SQS-100
structure at 8\% Mn concentration, as a function of Fermi level
(thus altering the filling of the $t_2$ level), i.e. additional
holes are modeled by simple Fermi level shifts.  The $J_{RR'}$ and
$T_c$ are maximal when the $t_2$ level contains approximately 3/2
electrons.

If this applies to both the $t_2$ and $e$ levels, the $t_2$ level
should contribute a FM exchange, while the $e$ level should
contribute a competing AFM exchange.  To see to what extent this
model describes the LDA exchange we resolve the mean-field estimate
for $T_c$ into separate $l$ and $m$ contributions, that is
\begin{eqnarray}
T^{\rm{}MFA}_c     &=& \sum_{L} T^{\rm{}MFA}_{c,L} \nonumber\\
T^{\rm{}MFA}_{c,L} &=& \frac{2}{3}\sum_{RR'L'}J_{RL'R'L'}
\end{eqnarray}
where $J_{RLR'L'}$ is identical to Eq.~(\ref{eq:jij}) but with the
sum over $LL'$ suppressed.  Three compounds with high Mn
concentration were studied: MnGa$_3$As$_4$ in the L1$_2$
structure, a four-atom SQS alloy Mn$_2$Ga$_2$As$_4$, and MnAs in
the ZB phase.  Data shown in Fig.~\ref{fig:jij} largely verifies
this picture.  The table in Fig.~\ref{fig:jij} shows the $e$
levels consistently contribute antiferromagnetically.  FM
contributions from $t_2$ level decrease with increasing Mn
concentration.  In the high concentration limit (ZB MnAs) the
picture of a $t_2$ derived impurity band is rather far removed
from reality.  Fig.~\ref{fig:jij} also shows site-resolved
$J_{RR'}$.  It shows that $J_{RR'}$ when $R$ points along [110]
tend to be FM, while those pointing along [001] tend to be AFM.
This tendency is compatible with symmetry of the $t_2$ orbitals,
which point in the [110] directions and are FM coupled, while the
$e$ orbitals point along the [001] are \emph{weak or even} AFM
coupled. Thus it seems that the LDA exchange interactions are
rather strongly dependent on symmetry of the Mn orbitals rather
than largely being a function of the Fermi surface, as RKKY-like
models suppose\cite{dietl00}.


The orientation dependence of $J$ has been studied previously by a
number of authors, mostly in the dilute case where the Mn doping
concentration is
$<10$\%\cite{Zhao04,Mahadevan_apl_04,Silva04,Bergqvist05,kudrnovsky04}.
One approach estimates $J$ from various FM-AFM
configurations\cite{mark01,Zhao04,Mahadevan_apl_04,Silva04}.
However, such calculations are predicated on the assumption that
large-angle rotations can map to a Heisenberg model (a
perturbation theory valid for small rotations).  More importantly,
this is approach is problematic especially for dilute case,
because many $J$'s are required.  The longer-ranged $J$s are
important to describe $T_c$, where percolation is important
\cite{Xu05}.  The more rigorously based linear-response
technique\cite{Bergqvist05} explicitly calculates energy changes
for small angle rotations.  Many $J$s are calculated at once; and
the range extends far beyond the unit cell.  In any case, none of
the preceding calculations yield AFM coupling in the [001]
direction, because the competing antiferromagnetic interactions
are much stronger at high concentrations\cite{mark01}, at least in
the LDA.
In the dilute case, we obtain both FM and AFM exchange coupling for [001]
direction, which when averaged yields a small net coupling, similar to a
CPA linear-response calculation\cite{Bergqvist05}. However, for high Mn
doping in the $\delta$-doped case studied here, AFM becomes increasingly
dominant along [001].

\begin{figure}[ht]
\centering
\epsfig{file=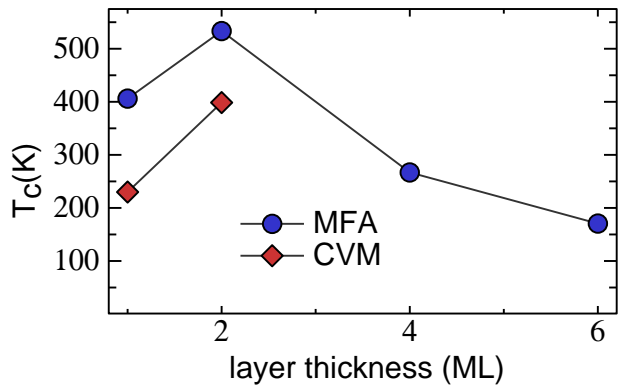,angle=0,width=0.45\textwidth,clip}
\caption{Dependence of $T_c$ on the $\delta$-layer thickness for
  an average Mn concentration of 50\% per layer.  Circles show
  mean-field results for comparison.}
\label{fig:thickness}
\end{figure}

The crystallographic dependence of $J_{RR'}$, and the reduction of
$J_{RR'}$ with Mn concentration, help to explain why $T_c$ is
maximal for partial fillings of the $\delta$ layer
(Fig.~\ref{fig:concentration}), and also suggest that $T_c$ will
be optimal for a rather small $\delta$ layer thickness.  To test
this, we compute $T_c$ varying the number $\delta$ layers.  We fix
the total thickness of the supercell as 8ML and change the
thickness of $\delta$-layer up to 8ML (corresponding to the bulk
case).  Each layer contains 50\% Mn and 50\% Ga (the 1-layer case
corresponds to the ``1/2'' point in Fig.~\ref{fig:concentration}).
Data is shown in Fig.~\ref{fig:thickness}.  $T_c$ is optimal at
2ML then decreases above that.  (For thicknesses $>$ 2 ML only
$T^{\rm{}MFA}_c$ is shown CVM predicts a spin glass).

\begin{figure}[ht]
\centering
\epsfig{file=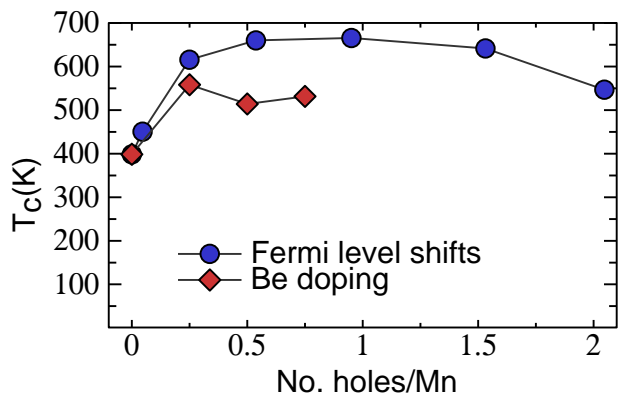,angle=0,width=0.45\textwidth,clip}
\caption{Dependence of $T_c$ on excess hole concentration for a
  2ML $\delta$-doped MnGaAs structure, with the $\delta$ layer
  containing 50\% Mn, as calculated by the CVM.  Hole doping is
  modeled through artificial Fermi level shifts, or induced
  through the addition of Be in GaAs layers adjacent to the
  $\delta$ layer.}
\label{fig:fermi}
\end{figure}

We now turn to the question: can $T_c$ be increased by co-doping
to add holes?  We can mimic addition of holes by Fermi level
shifts (as before); alternatively we can dope the GaAs layer with
an acceptor such as Be.  In Ref.~\onlinecite{Nazmul05}, $\delta$-doped
MnGaAs were grown in this way.  Extra holes were supplied to the
$\delta$-layer by doping Be in the GaAs near (but not in) the
$\delta$-layer, to prevent compensating donors such as Mn
interstitials from forming in the $\delta$-layer.

In Fig.~\ref{fig:fermi}, the dependence of $T_c$ on the number of extra
holes is presented.  $T_c$ increases markedly by the addition of $\sim 1/2$
hole/Mn atom.  Higher hole concentrations do not further increase $T_c$ (as
in the random alloy).  The difference between Fermi level shifts and using
actual dopants can be explained in that the latter contains electrostatic
shifts as a result of charge transfer from the Be dopants to the Mn layers,
thus reducing hole confinement in the $\delta$ layer. This confirms the
experimental results for $\delta$ layer and also suggest a much higher
$T_c$ for defect free samples compared with the best $T_c$ recorded to
date.

In conclusion, we demonstrate a strong crystallographic
dependence of magnetic exchange interactions in {\gamnas}, and
show that it can be exploited in $\delta$-doped Mn:GaAs
structures to optimize $T_c$.  We determine the size and
concentration of $\delta$-layers, and hole concentration that
are predicted to result in an optimal $T_c$.  Under optimal
conditions (which entails superlattices to be grown sufficiently
defect-free) $T_c$ is predicted to be well above room
temperature.

This work was supported by ONR contract N00014-02-1-1025.


\begin{thebibliography}{30}
\bibitem[{Two()}]{TwoDHeisennote}
\bibinfo{note}{It was recently shown by an analysis of the 2D Heisenberg
  ferromagnet within the RPA, that even though the vanilla Heisenberg
  hamiltonian orders at 0K\cite{mermin66}, a small amount anisotropy acts as a
  ``trigger'' to stabilize $T_c$ ($T_c$ depends weakly on the anisotropy when
  it is small)\cite{bruno06}. We compute the anisotropy for some 1 ML cases, and find that it
  depends rather sensitively on the detailed configuration of atoms in the
  monolayer, ranging from ~0.3 to 20 meV. It sufficient to stabilize the FM in
  this 2D structure, so we can reasonably neglect it in these calculations.}
\end{thebibliography}

\end{document}